\documentclass[conference]{IEEEtran}
\IEEEoverridecommandlockouts

\usepackage{cite}
\usepackage{amsmath,amssymb,amsfonts}
\usepackage{algorithmic}
\usepackage{graphicx}
\usepackage{textcomp}

\newcommand{\ceq}{\mathop{=}\limits}

\newcommand{\R}{\mathbb{R}}
\newcommand{\C}{\mathbb{C}}
\newcommand{\Rp}{\R_{\geq 0}}

\newcommand{\vecf}[1]{\boldsymbol{#1}}
\newcommand{\matf}[1]{\boldsymbol{#1}}
\newcommand{\transpose}{\mathsf{T}}
\newcommand{\Hermitian}{\mathsf{H}}

\renewcommand{\log}{\ln}

\newcommand{\DNN}{\mathrm{DNN}}

\def\thline{\noalign{\hrule height 1.0pt}}
\def\tthline{\noalign{\hrule height 1.4pt}}

\allowdisplaybreaks
\def\mysection#1{\vspace{-1mm}\section{#1}\vspace{-1mm}}
\def\mysubsection#1{\vspace{-1mm}\subsection{#1}\vspace{-1mm}}

\def\BibTeX{{\rm B\kern-.05em{\sc i\kern-.025em b}\kern-.08em
    T\kern-.1667em\lower.7ex\hbox{E}\kern-.125emX}}

\begin{document}

\title{Empirical Bayesian Independent Deeply Learned Matrix Analysis For Multichannel Audio Source Separation
\thanks{This work was supported by JSPS-CAS Joint Research Program, Grant number JPJSBP120197203, and JSPS KAKENHI Grant Numbers JP19K20306, JP19H01116, and JP17H06101.}
}

\author{
    \IEEEauthorblockN{
        Takuya Hasumi$^{\dagger}$,
        Tomohiko Nakamura$^{\dagger}$,
        Norihiro Takamune$^{\dagger}$ \\
        Hiroshi Saruwatari$^{\dagger}$,
        Daichi Kitamura$^{\star}$,
        Yu Takahashi$^{\ddagger}$,
        Kazunobu Kondo$^{\ddagger}$
    }
    \IEEEauthorblockA{
        \textit{$^{\dagger}$The University of Tokyo, Tokyo, Japan} \\
        \textit{$^{\star}$National Institute of Technology, Kagawa College, Kagawa, Japan} \\
        \textit{$^{\ddagger}$Yamaha Corporation, Shizuoka, Japan} \\
    }
}

\maketitle

\begin{abstract}
Independent deeply learned matrix analysis (IDLMA) is one of the state-of-the-art supervised multichannel audio source separation methods.
It blindly estimates the demixing filters on the basis of source independence, using the source model estimated by the deep neural network (DNN).
However, since the ratios of the source to interferer signals vary widely among time-frequency (TF) slots, it is difficult to obtain reliable estimated power spectrograms of sources at all TF slots.
In this paper, we propose an IDLMA extension, empirical Bayesian IDLMA (EB-IDLMA), by introducing a prior distribution of source power spectrograms and treating the source power spectrograms as latent random variables.
This treatment allows us to implicitly consider the reliability of the estimated source power spectrograms for the estimation of demixing filters through the hyperparameters of the prior distribution estimated by the DNN.
Experimental evaluations show the effectiveness of EB-IDLMA and the importance of introducing the reliability of the estimated source power spectrograms.
\end{abstract}


\mysection{Introduction}
\label{sec:introduction}
Multichannel audio source separation aims at separating individual sources from a multichannel mixture signal observed using a microphone array \cite{Sawada2019APSIPATSIP}.
In an overdetermined or determined case, where the number of microphones is greater than or equal to that of sources, many blind source separation (BSS) methods based on the statistical source independence have been proposed for decades, for example, independent component analysis~\cite{comon1994independent} and independent vector analysis~\cite{kim2006blind,hiroe2006solution}. 
One of the state-of-the-art BSS methods is independent low-rank matrix analysis (ILRMA)~\cite{kitamura2016ASLP}, which estimates demixing filters using a non-negative matrix factorization (NMF)~\cite{lee1999learning} as the source model.
To elaborate the source model, by introducing deep neural networks (DNNs) into the ILRMA framework, we previously proposed independent deeply learned matrix analysis (IDLMA)~\cite{8747523}, one of the state-of-the-art supervised methods.

Since the demixing filters of IDLMA are updated using the power spectrograms of sources estimated by the DNNs, the separation performance depends on the estimation accuracy of the power spectrograms.
However, the ratios of the source to interferer signals vary widely among time-frequency (TF) slots; thus, it is generally difficult to obtain reliable estimated power spectrograms at all TF slots.
Nevertheless, the DNNs of IDLMA are designed to estimate only the power spectrograms but not their reliability measures.
Owing to the lack of these measures, we have no choice but to use the estimated power spectrograms despite that they may fail in source separation at some TF slots.

In this paper, we extend IDLMA to allow the demixing filter estimation while taking into account the reliability of the source power spectrograms obtained with the DNNs.
We introduce the prior distribution of the source model into the IDLMA model and treat the source power spectrograms as latent random variables.
By marginalizing them out, we can implicitly consider the reliability of the estimated source power spectrograms for the demixing filter estimation through the hyperparameters of the prior distribution.
We train the DNNs to estimate the hyperparameters that maximize the marginal likelihood of the observed signals.
This hyperparameter estimation method is called an empirical Bayes method, and we call the proposed extension the empirical Bayesian IDLMA (EB-IDLMA).

\mysection{Independent Deeply Learned Matrix Analysis \cite{8747523}}
\label{sec:conventional-method}
\mysubsection{Formulation}
\label{sec:conventional-method/formulation}
\begin{figure*}[t]
    \centering
    \includegraphics[width=1.6\columnwidth,clip]{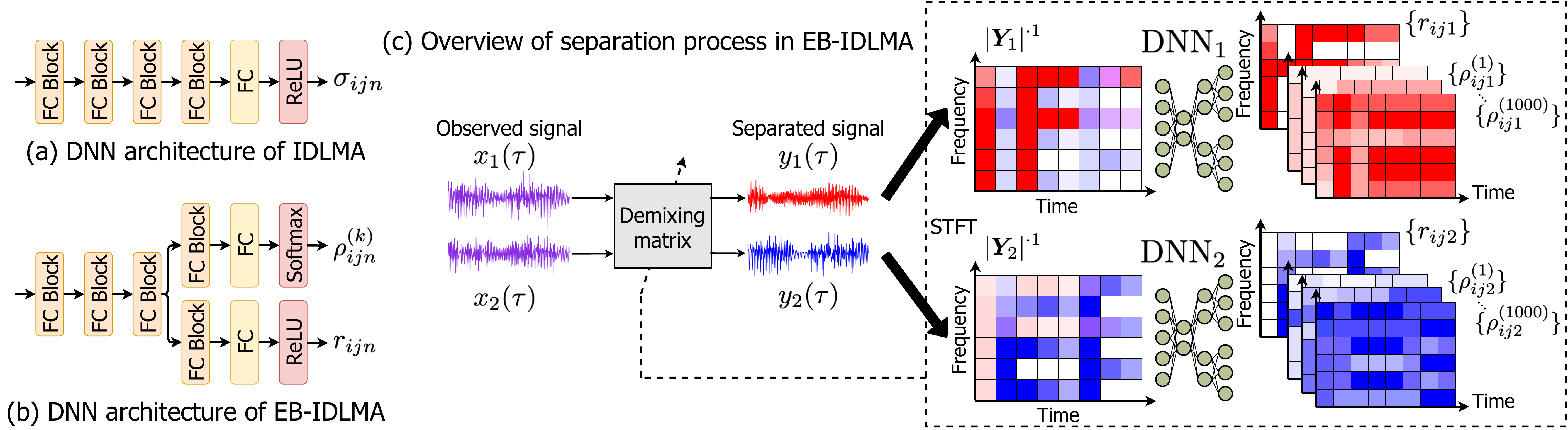}
    \caption{DNN architectures of (a) IDLMA and (b) EB-IDLMA, and (c) overview of separation process in proposed EB-IDLMA.}
    \label{fig:outline}
\end{figure*}
Let us respectively denote the numbers of sources and channels by $N$ and $M$.
The short-time Fourier transforms (STFTs) of source, observed, and separated signals are respectively given by
$\vecf{s}_{ij} = (s_{ij1}, \ldots, s_{ijN})^{\transpose}\in\mathbb{C}^{N}$,
$\vecf{x}_{ij} = (x_{ij1}, \ldots, x_{ijM})^{\transpose}\in \mathbb{C}^{M}$,
and $\vecf{y}_{ij} = (y_{ij1}, \ldots, y_{ijN})^{\transpose}\in \mathbb{C}^{N}$,
where $i=1,\ldots,I,j=1,\ldots,J,n=1,\ldots,N,$ and $m=1,\ldots,M$ are the frequency, frame, source, and channel indices, respectively, and ${ }^{\transpose}$ denotes the transpose operator.
We also denote a matrix whose $(i,j)$th entry is $x_{ijm}$ ($y_{ijn}$) as $\matf{X}_{m}\in\C^{I\times J}$ ($\matf{Y}_{n}\in\C^{I\times J}$).
When the mixing system is linear time-invariant and the analysis window of the STFT is sufficiently longer than the room impulse response, the mixing system $\matf{A}_{i}\in\C^{M\times N}$ is instantaneous:
$\vecf{x}_{ij} = \matf{A}_{i}\vecf{s}_{ij}.$
When $M=N$ and $\matf{A}_i$ is a nonsingular matrix, the estimated signals can be computed using a demixing matrix $\matf{W}_{i}=(\vecf{w}_{i1},\ldots,\vecf{w}_{iN})^{\mathsf{H}}$, where ${ }^{\Hermitian}$ denotes the Hermite transpose:
\begin{equation}
    \vecf{y}_{ij} = \matf{W}_{i}\vecf{x}_{ij}.
    \label{eq:demixing}
\end{equation}

IDLMA adopts the so-called local Gaussian model, i.e., $y_{ijn}$ is assumed to be conditionally independent w.r.t. $i$ and $j$, and follow an isotropic complex Gaussian distribution:
\begin{equation}
    p(y_{ijn}|\sigma^2_{ijn})
    = \dfrac{1}{\pi\sigma_{ijn}^2}\exp{\left(-\dfrac{|y_{ijn}|^2}{\sigma_{ijn}^2}\right)},
    \label{eq:lgm}
\end{equation}
where $\sigma_{ijn}$ is the scale parameter.
Using the change-of-variable technique, we can describe the cost function of IDLMA as the negative log-likelihood of $\mathcal{X}=\{\matf{X}_{1},\ldots,\matf{X}_{M}\}$~\cite{Sawada2019APSIPATSIP}:
\begin{align}
    \mathcal{L}_{\mathrm{Gauss}}
    = &-\log p(\mathcal{X}) \notag \\
    = &-\log p(\mathcal{Y}) - 2J\sum_{i}\log|\det\matf{W}_{i}| \notag \\
    \ceq^{c}&\sum_{i,j,n}
    \left(\log\sigma_{ijn}^{2} + \frac{|\vecf{w}_{in}^{\Hermitian}\vecf{x}_{ij}|^{2}}{\sigma_{ijn}^{2}}\right) - 2J\sum_{i}\log|\det\matf{W}_{i}|,
    \label{eq:L_gauss}
\end{align}
where $\mathcal{Y} =\{\matf{Y}_{1},\cdots,\matf{Y}_{N}\}$ is the set of estimated signals and $\ceq^{c}$ denotes the equality up to constants.
The minimization of $\mathcal{L}_{\mathrm{Gauss}}$ w.r.t. $\matf{W}_i$ amounts to the maximization of the statistical independence between the sources.

\mysubsection{DNN Training}
\label{sec:conventional-method/training-procedure}
We first train the DNN of source $n$, say $\DNN_{n}$, to estimate the magnitude spectrogram of the target source from that of the single-channel noisy mixture.
Let us denote the $(i,j)$th entry of the groundtruth complex spectrogram of source $n$ by $\tilde{s}_{ijn}\in\C$ and that of the estimated magnitude spectrogram by $\hat{\sigma}_{ijn}\in\Rp$.
The cost function of the DNN training is defined by the Itakura-Saito divergence as follows:
\begin{equation}
    \mathcal{L}_{\mathrm{Gauss}}^{(\DNN_{n})}
    = \sum_{i,j}
    \left(\frac{|\tilde{s}_{ijn}|^{2}+\delta}{\hat{\sigma}_{ijn}^{2}+\delta}
    - \log\frac{|\tilde{s}_{ijn}|^{2}+\delta}{\hat{\sigma}_{ijn}^{2}+\delta}- 1\right),
    \label{eq:L_gauss_dnn}
\end{equation}
where $\delta$ is a small value to avoid division by zero.
Since the replacement of $|\vecf{w}_{in}^{\Hermitian}\vecf{x}_{ij}|$ with $|\tilde{\vecf{s}}_{ijn}|$ reduces the first term of \eqref{eq:L_gauss} to \eqref{eq:L_gauss_dnn} up to $\delta$, the minimization of \eqref{eq:L_gauss_dnn} with $\hat{\sigma}_{ijn}$ can be interpreted as a simulation of the maximum likelihood estimation of $\sigma_{ijn}$ based on \eqref{eq:L_gauss_dnn}.
This interpretation approximately justifies the use of the trained DNNs as the source models for the following demixing matrix estimation.

\mysubsection{Demixing Matrix Estimation}
\label{sec:conventional-method/idlma}
After the DNN training, IDLMA estimates $\matf{W}_i$ from the observed signal $\vecf{x}_{ij}$ without {\em a priori} spatial information by iteratively performing the following two steps:
(i) The demixing matrix $\matf{W}_i$ is updated according to an efficient and convergence-guaranteed optimization algorithm, the iterative projection (IP) algorithm~\cite{ono2011stable}, which can be applied to the sum of a negative log-determinant and a quadratic form.
Owing to space limitation, we omitted the details of the IP algorithm (see \cite{ono2011stable} for details).
(ii) The scale parameter $\sigma_{ijn}$ is updated using the pretrained DNNs as
\begin{equation}
    \sigma_{ijn} \leftarrow \max([\DNN_{n}(|\matf{Y}_{n}|^{\cdot 1})]_{ij},\varepsilon),
\end{equation}
where $|\cdot|^{\cdot 1}$ returns the elementwise absolute values, $[\cdot]_{ij}$ returns the $(i,j)$th entry of a matrix, and $\varepsilon$ is a small value for numerical stability.

\mysection{PROPOSED METHOD}
\mysubsection{Introduction of Prior Distribution}
Although $\sigma_{ijn}^2$ is treated as the parameter in IDLMA, we treat $\sigma_{ijn}^2$ in the proposed EB-IDLMA as a latent random variable.
When an appropriate prior distribution is introduced, this treatment enables us to derive a posterior distribution of $\sigma_{ijn}^2$ and further marginalize $\sigma_{ijn}^2$ out.
The marginalization allows us to implicitly consider the reliability of $\sigma_{ijn}^2$ using hyperparameters of the prior distribution.

Let us assume that $\sigma_{ijn}^2$ follows the inverse gamma distribution:
\begin{equation}
    p(\sigma_{ijn}^{2};a_{ijn},b_{ijn}) 
    = \cfrac{b_{ijn}^{a_{ijn}}}{\Gamma(a_{ijn})}{\left(\cfrac{1}{\sigma_{ijn}^{2}}\right)}^{{a_{ijn}+1}}\exp{\left(-\cfrac{b_{ijn}}{\sigma_{ijn}^{2}}\right)},
    \label{eq:prior}
\end{equation}
where $a_{ijn}>0$ and $b_{ijn}>0$ are the shape and scale parameters, respectively.
Since the inverse gamma distribution is the conjugate prior of the Gaussian distribution, $\sigma_{ijn}$ can be marginalized out, and the marginal likelihood $p(y_{ijn};a_{ijn},b_{ijn})$ is given as
\begin{align}
    p(y_{ijn};a_{ijn},b_{ijn}) 
    =& \frac{a_{ijn}b_{ijn}^{a_{ijn}}}{\pi(|y_{ijn}|^2+b_{ijn})^{a_{ijn}+1}}.
    \label{eq:marginal}
\end{align}
This probability distribution is indeed the complex Student's-$t$ distribution with the scale parameter of $r_{ijn}^2:=b_{ijn}/a_{ijn}$ and the degree of freedom parameter of $\nu_{ijn}:=2a_{ijn}$~\cite{bishop2006pattern}, which determines the reliability of $r_{ijn}$ as shown later in Section~\ref{sec:demixing}.
By using $\nu_{ijn}$ and $r_{ijn}^{2}$, we can write the cost function of EB-IDLMA  as the negative marginal log-likelihood:
\begin{align}
    \mathcal{L}_{\mathrm{EB}}
    \ceq^{c} & \sum_{i,j,n}\left[\log r_{ijn}^{2} + \left(1+\frac{\nu_{ijn}}{2}\right)\log\left(1+\frac{2|\vecf{w}_{in}^{\Hermitian}\vecf{x}_{ij}|^{2}}{\nu_{ijn}r_{ijn}^{2}}\right)\right] \nonumber \\
    &- 2J\sum_{i}\log|\det\matf{W}_{i}|.
    \label{eq:negative_log-likelihhod_EB-IDLMA}
\end{align}

\mysubsection{DNN Training Based on Empirical Bayes Method}
The key concept of IDLMA is to use a cost function consistent with that of the demixing matrix estimation for training the DNNs.
To maintain this concept, we train the DNNs on the basis of the empirical Bayes method since the problem of minimizing $\mathcal{L}_{\mathrm{EB}}$ w.r.t. $r_{ijn}^2$ and $\nu_{ijn}$ is equivalent to that of maximizing the marginal likelihood w.r.t. the hyperparameters of the prior distribution.

Let $\hat{\nu}_{ijn}$ and $\hat{r}_{ijn}$ be the hyperparameters estimated by the DNN of source $n$ at the $(i,j)$th TF slot.
Similary to IDLMA (see Section \ref{sec:conventional-method/training-procedure}), we define the cost function for the DNN training to be consistent with $\mathcal{L}_{\mathrm{EB}}$.
\begin{align}
    \mathcal{L}_{\mathrm{EB}}^{(\DNN_{n})}
    =& \sum_{i,j}\log(\hat{r}_{ijn}^{2} + \delta) \nonumber \\
    & + \sum_{i,j}\left(1+\frac{\hat{\nu}_{ijn}}{2}\right)\log\left[1+
    \frac{2(|\tilde{s}_{ijn}|^{2}+\delta)}{\hat{\nu}_{ijn}(\hat{r}_{ijn}^{2}+\delta)}
    \right].
    \label{eq:NLL_EB-IDLMA_DNN}
\end{align}

Although the DNNs can be designed to directly output $\hat{r}_{ijn}$ and $\hat{\nu}_{ijn}$, we experimentally found that the estimates of $\hat{\nu}_{ijn}$ extremely increased at the TF slots with near-zero energy during training.
This observation can also be confirmed theoretically by the following proposition:

\noindent \textbf{Proposition~1}  When $|\tilde{s}_{ijn}|^2\ll \delta$ and $\hat{r}_{ijn}^2\ll \delta$, the cost function $\mathcal{L}_{\mathrm{EB}}^{(\DNN_{n})}$ approximately decreases as $\hat{\nu}_{ijn}$ increases.

\noindent The proof of Proposition~1 is shown in Appendix~\ref{sec:appendix/proof1}.
These results show that the DNNs are likely to be obsessed with increasing $\nu_{ijn}$ at the TF slots with near-zero energy, which may be one of the causes of performance degradation.

One method to prevent the excessive increase in $\hat{\nu}_{ijn}$ is to represent $\hat{\nu}_{ijn}$ as a weighted sum of a limited number of anchors:
\begin{equation}
    \hat{\nu}_{ijn}
    = \sum_{k\in\mathcal{K}}\rho_{ijn}^{(k)}k,
    \label{eq:categorical}
\end{equation}
where $\mathcal{K}$ is the set of anchors and $\rho_{ijn}^{(k)}$ is the weight of anchor $k$ such that $0\leq \rho_{ijn}^{(k)}\leq 1$ and $\sum_{k}\rho_{ijn}^{(k)}=1$ for all $i$, $j$, and $n$.
This representation allows us to restrict the $\hat{\nu}_{ijn}$ value within a range from the minimum anchor to the maximum one.
To encompass the $\hat{\nu}_{ijn}$ representation, we design the DNN to output $\rho_{ijn}^{(k)}$ instead of $\hat{\nu}_{ijn}$.
Note that although we can restrict the range of $\hat{\nu}_{ijn}$ by clipping its value, we experimentally found that the representation given by \eqref{eq:categorical} achieved the higher performance.

Figs.~\ref{fig:outline}(a) and (b) show the DNN architectures of the conventional IDLMA and proposed EB-IDLMA, respectively.
In the conventional IDLMA, the DNN of source $n$ outputs only the scale parameters and consists of five fully connected (FC) blocks, each of which is composed of a FC layer with $2048$ hidden units, rectified linear unit (ReLU) nonlinearity, and a dropout layer with a drop rate of $0.3$.
On the other hand, the DNN of EB-IDLMA outputs $\rho_{ijn}^{(k)}$s and $\hat{r}_{ijn}$s.
It is a two-headed network consisting of three FC blocks followed by two separate head subnetworks for $\rho_{ijn}^{(k)}$ and $\hat{r}_{ijn}$.
Each subnetwork is composed of two FC blocks, but the ReLU nonlinearity in the last FC block of the subnetwork for $\hat{\nu}_{ijn}$ is replaced with softmax nonlinearity.
In all DNNs, the dropout layers of the last FC blocks are removed.

\mysubsection{Demixing Matrix Estimation} \label{sec:demixing}
Given the STFTs of the observed signals $\vecf{x}_{ij}$, we seek to find the demixing matrix $\matf{W}_i$ that minimizes the cost function of EB-IDLMA $\mathcal{L}_{\mathrm{EB}}$, where $\nu_{ijn}$ and $r_{ijn}$ are computed with the pretrained DNNs as
\begin{equation}
    \nu_{ijn}\leftarrow \sum_{k\in\mathcal{K}}\rho_{ijn}^{(k)}k, \quad
    r_{ijn}\leftarrow \max (\hat{r}_{ijn}, \varepsilon).
    \label{eq:update_nu_r}
\end{equation}
Unlike IDLMA, the IP algorithm cannot be directly applied to this minimization problem since $\mathcal{L}_{\mathrm{EB}}$ includes $|\vecf{w}_{in}^{\Hermitian}\vecf{x}_{ij}|$ in the logarithm function.
However, we can transform $\mathcal{L}_{\mathrm{EB}}$ into an IP-applicable form, i.e., the sum of a negative log-determinant and a quadratic form, in the same manner as in \cite{8747523}, where the majorization--minimization (MM) algorithm~\cite{hunter2000quantile} is employed.

In the MM algorithm, by introducing auxiliary variables, we construct an upper bound of $\mathcal{L}_{\mathrm{EB}}$ (majorization function) that equals $\mathcal{L}_{\mathrm{EB}}$ at exactly one point.
We then alternately update the original parameters and auxiliary variables so that they minimize the majorization function, which guarantees the non-increase in $\mathcal{L}_{\mathrm{EB}}$.

Focusing on the fact that a logarithm function is lower than or equal to its tangent line owing to its concavity, we can derive the upper bound of the logarithm term including $|\vecf{w}_{in}^{\Hermitian}\vecf{x}_{ij}|$ of \eqref{eq:negative_log-likelihhod_EB-IDLMA} as
\begin{eqnarray}
    \lefteqn{\log\left(1{+}\frac{2|\vecf{w}_{in}^{\Hermitian}\vecf{x}_{ij}|^{2}}{\nu_{ijn}r_{ijn}^2}\right)} \notag \\
    & \leq &\cfrac{1}{\gamma_{ijn}}\left(
        1{+}\cfrac{2|\vecf{w}_{in}^{\Hermitian}\vecf{x}_{ij}|^{2}}{\nu_{ijn}r_{ijn}^2}-\gamma_{ijn}
    \right) {+} \log \gamma_{ijn},
    \label{eq:auxfun}
\end{eqnarray}
where $\gamma_{ijn}$ is an auxiliary variable.
The equality holds if and only if $\gamma_{ijn}=1+2|\vecf{w}_{in}^{\Hermitian}\vecf{x}_{ij}|^{2}/(\nu_{ijn}r_{ijn}^2)$.
By using inequality \eqref{eq:auxfun}, we can obtain the upper bound of $\mathcal{L}_{\mathrm{EB}}$ as
\begin{align}
    \mathcal{L}_{\mathrm{EB}}^{+} \ceq^{c}
    & J\sum_{i,n}\vecf{w}_{in}^{\Hermitian}\matf{U}_{in}\vecf{w}_{in} -2J \sum_{i}\log|\det\matf{W}_{i}|,
    \label{eq:aux_L_EB} \\
    \matf{U}_{in} :=&
    \frac{1}{J}\left(1+\frac{\nu_{ijn}}{2}\right)\sum_{j}\frac{1}{\gamma_{ijn}\sigma_{ijn}^{2}}\vecf{x}_{ijn}\vecf{x}_{ijn}^{\Hermitian},
        \label{eq:Uin_EB-IDLMA}
\end{align}
where we only show the terms including the demixing filters.
Substituting the equality condition of inequality \eqref{eq:auxfun} into \eqref{eq:Uin_EB-IDLMA} yields 
\begin{align}
    \matf{U}_{in}
    =& \frac{1}{J}\sum_{j}\frac{1}{\xi_{ijn}}\vecf{x}_{ij}\vecf{x}_{ij}^{\Hermitian},
    \label{eq:Uin_readable}
    \\
    \xi_{ijn}
    =& \cfrac{\nu_{ijn}}{\nu_{ijn}+2}\,r^2_{ijn} + \cfrac{2}{\nu_{ijn}+2}\,|y_{ijn}|^{2}.
    \label{eq:xi}
\end{align}
Since the right-hand side of \eqref{eq:aux_L_EB} is the sum of a negative log-determinant and a quadratic form, we can update $\matf{W}_i$ by the IP algorithm, after which we update the separated signals as $y_{ijn}\leftarrow \vecf{w}_{in}^{\Hermitian}\vecf{x}_{ij}$ and apply the back-projection technique to it to compensate for the scale indeterminacy \cite{Murata2001NC}.
To sum up, the separation algorithm of EB-IDLMA is to iteratively perform the update of $\matf{W}_i$ according to the IP algorithm and that of $\nu_{ijn}$ and $r_{ijn}$ according to the update rule \eqref{eq:update_nu_r}, as shown in Fig.~\ref{fig:outline}(c).

Since $\xi_{ijn}$ is a convex combination of the estimated source power $r_{ijn}^2$ and the power of the separated signal $|y_{ijn}|^2$, we can interpret $\nu_{ijn}/(\nu_{ijn}+2)$ as the reliability of $r_{ijn}^2$.
The larger $\nu_{ijn}$ is the more the update of $\matf{W}_i$ depends on $r_{ijn}^2$.

\mysubsection{Relationship with Prior Works}
The proposed EB-IDLMA approximately reduces to IDLMA as $\nu_{ijn}\rightarrow \infty$, which results in the reliability of $r_{ijn}$ not being considered in the update of $\matf{W}_i$ since $\xi_{ijn}\simeq r_{ijn}$.
If $\nu_{ijn}$ is predetermined at the same value for all $i$, $j$, and $n$, denoted by $\nu$, EB-IDLMA reduces to $t$-IDLMA \cite{8747523}, where $y_{ijn}$ is assumed to follow an isotropic complex Student's-$t$ distribution.
Since the $\nu$ value is predetermined, the DNNs of $t$-IDLMA must be trained for each $\nu$ value.
However, we have no information on the best $\nu$ to choose before the separation, and the separation performance of $t$-IDLMA greatly depends on the $\nu$ value, as shown later in Section~\ref{sec:experimental-evaluation/result}.
Thus, the exploration of the best $\nu$ is computationally expensive, whereas our proposed EB-IDLMA does not require such exploration.
Furthermore, since $t$-IDLMA uses the same $\nu$ value at all TF slots, it cannot deal with the difference in the reliability of $r_{ijn}$ among the TF slots.

\mysection{EXPERIMENTAL EVALUATION}
\label{sec:experimental-evaluation}

\mysubsection{Experimental Setting}
\label{sec:experimental-evaluation/setting}

\begin{figure}[t]
\centering
\centerline{\includegraphics[width=0.8\columnwidth]{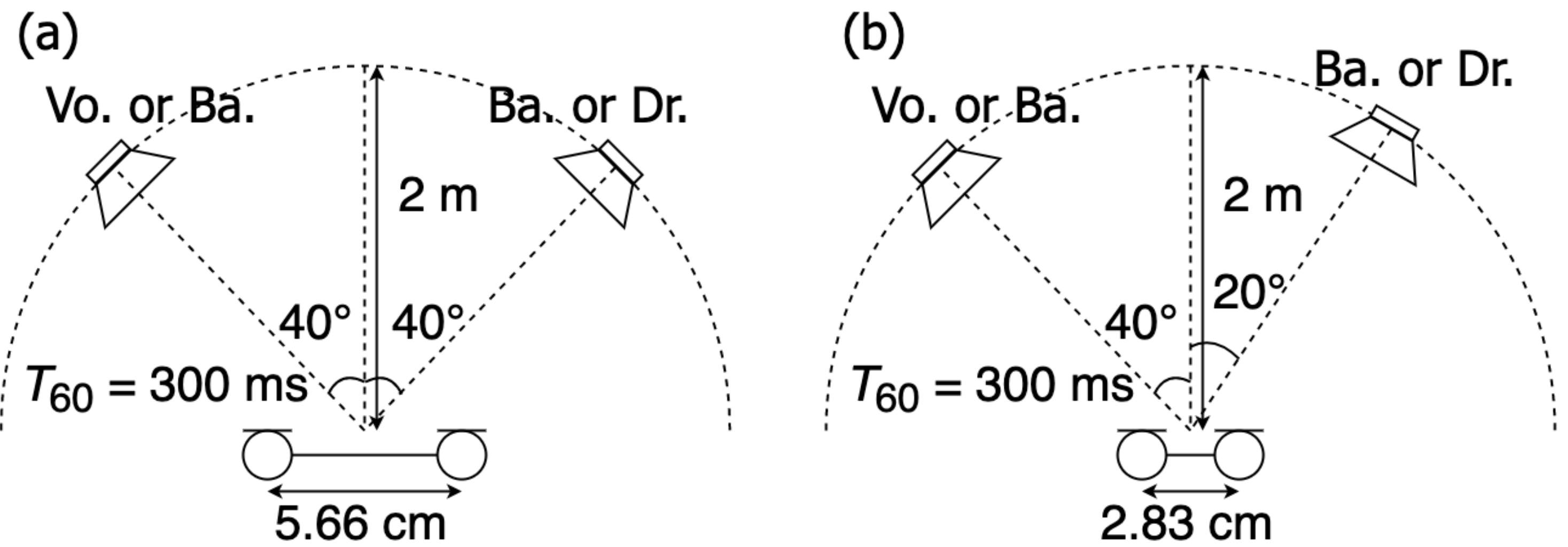}}
\caption{Recording conditions.}
\label{fig:experimental-evaluation/recording-condition}
\end{figure}
\begin{figure}[t]
    \centering
    \includegraphics[width=0.9\columnwidth]{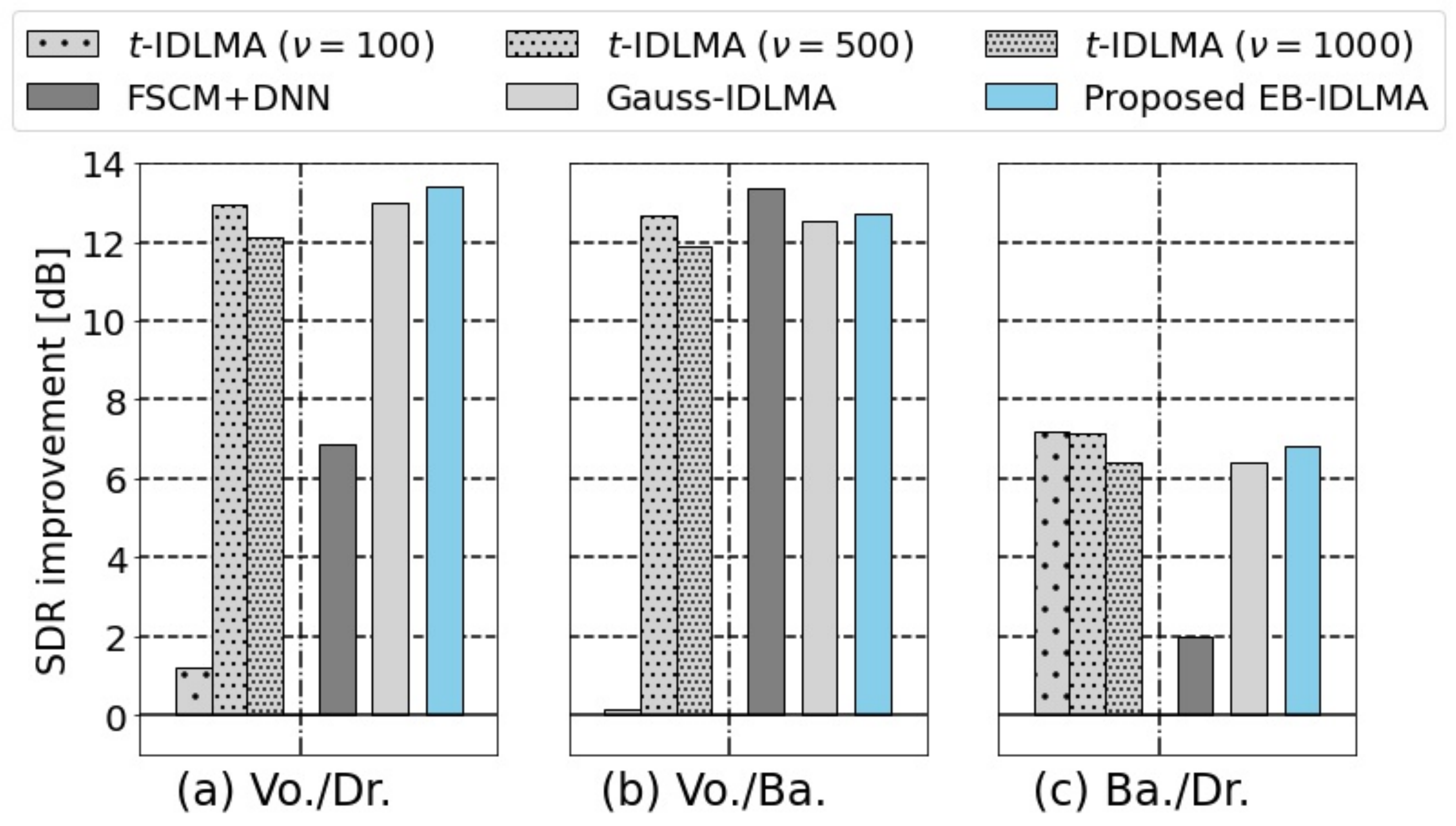}
    \caption{SDRs of conventional and proposed methods.}
    \label{fig:experimental-evaluation/result}
\end{figure}
\begin{figure}[t]
    \centering
    \includegraphics[width=1\columnwidth]{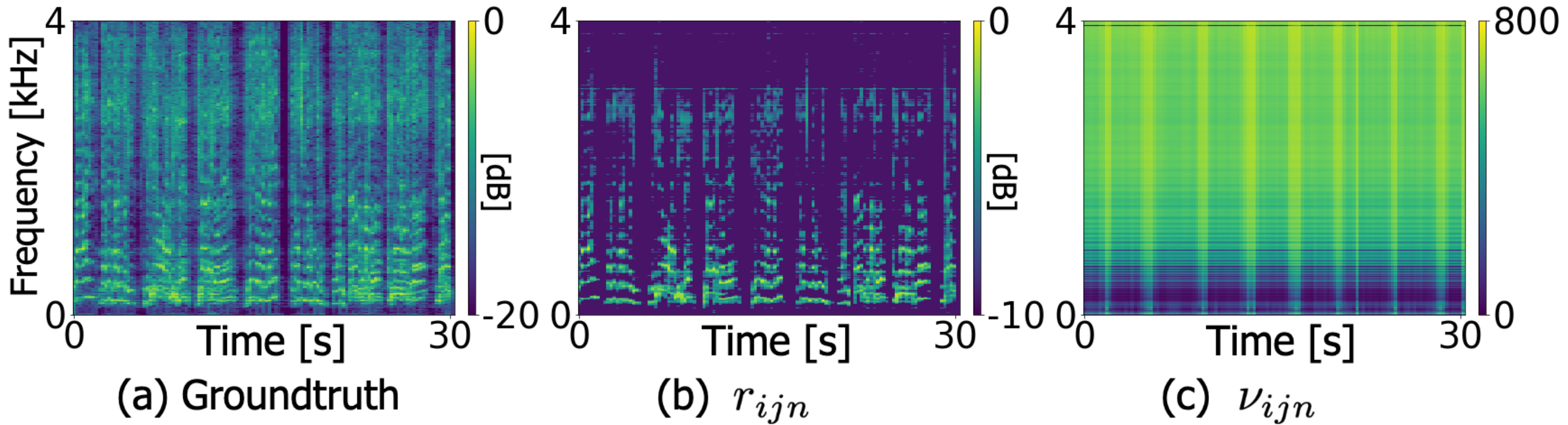}
    \caption{Examples of groundtruth spectrogram, $r_{ijn}$, and $\nu_{ijn}$ of vocal obtained with DNN of EB-IDLMA for Vo./Ba. separation.}
    \label{fig:experimental-evaluation/DoF}
\end{figure}
To evaluate the effectiveness of EB-IDLMA, we conducted multichannel audio source separation experiments using the DSD100 dataset~\cite{liutkus20172016}.
We used a $512$-ms hamming window with a $256$-ms hopsize for STFT and downsampled all audio signals to $8$ kHz.
For the test data, we used the $30$- to $60$-s segments of the top $25$ songs in alphabetical order in the \texttt{test} set as dry sources.
By convolving these sources with the impulse responses ($T_{60}=300$ ms) of the RWCP database~\cite{nakamura2000acoustical}, whose recording conditions are shown in Fig.~\ref{fig:experimental-evaluation/recording-condition}, we created in total $50$ two-channel mixtures of each pair of three musical instruments: vocal (Vo.), bass (Ba.), and drums (Dr.).

We compared EB-IDLMA with the combination of the full-rank spatial covariance model with DNN (FSCM+DNN)~\cite{nugraha2016multichannel}, and IDLMA (Gauss-IDLMA).
For all models, we updated the spatial model $100$ times and the source power spectrograms by the DNNs every $10$ iterations of the spatial model update.
For a fair comparison, we used the same DNNs for FSCM+DNN as Gauss-IDLMA.

We trained the DNNs using the \texttt{dev} set ($50$ songs) for training and the bottom $25$ songs in alphabetical order in the \texttt{test} set for validation.
Note that the conventional and proposed methods are spatially blind, and the audio signals used in the DNN training were not convolved with any room impulse response.
The batch size was set at $128$.
The minibatch generation procedure was the same as those in \cite{8747523} except for random gains for the sources.
We randomly generated the gains from a uniform distribution over $[0.05,1]$ (a beta distribution with shape parameters of $0.1$ and $1$) for the target source (interferer).
This gain generation simulated that $\matf{Y}_n$ tend to include less interferers in the latter iterations of the spatial model update.
The DNNs of the conventional and proposed methods were trained using the Adadelta optimizer~\cite{zeiler2012adadelta} with a weight decay of $10^{-5}$ for $2000$ epochs.
Gradient clipping with the maximum norm of $10$ was applied.
The other hyperparameters were set as $\mathcal{K} = \{1,10,100,1000\}$, $c=3$, $\delta=10^{-5}$, and $\varepsilon=10^{-1/2}$.

\mysubsection{Results}
\vspace{-0.6mm}
\label{sec:experimental-evaluation/result}
\begin{table}[t]
    \centering
    \caption{Ratios of $\nu$ with which $t$-IDLMA gave highest performance}
    \begin{tabular}{c|ccc} \tthline
         Instrument pair   & $\nu=100$ & $\nu=500$ & $\nu=1000$ \\ \thline
        Vo./Dr. & $0$ \% & $\mathbf{64}$ \% & $36$ \% \\ 
        Vo./Ba. & $0$ \% & $46$ \% & $\mathbf{54}$ \% \\
        Ba./Dr. & $\mathbf{42}$ \% & $34$ \% & $24$ \% \\
        \tthline
    \end{tabular}
    \label{tab:ratio_nu_tIDLMA}
\end{table}
Fig.~\ref{fig:experimental-evaluation/result} shows signal-to-distortion ratio (SDR) improvements of all methods, which were computed using the BSSEval toolbox~\cite{vincent2006performance} and averaged over the $50$ test mixtures for each instrument pair.
Although the SDR improvements of FSCM+DNN greatly varied with the instrument pairs, Gauss- and EB-IDLMAs worked robustly against the instrument variations.
The proposed EB-IDLMA provided higher SDR improvements than Gauss-IDLMA for all instrument pairs, showing the effectiveness of considering the reliability of the estimated source model.
As shown in Fig.~\ref{fig:experimental-evaluation/DoF}, the estimated degree of freedom parameter $\nu_{ijn}$ varied with frequency, which means that the estimated source model $r_{ijn}$ was less reliable in the lower frequency band.
This should be due to the fact that the spectrograms of Ba. and Vo. markedly overlapped in the lower frequency band.

Fig.~\ref{fig:experimental-evaluation/result} also shows the separation results of $t$-IDLMA with $\nu=100,500$, and $1000$.
The SDR improvements of $t$-IDLMA greatly varied with $\nu$, demonstrating that $\nu$ should be carefully predetermined.
Compared with $t$-IDLMA using the best $\nu$ for each instrument pair, EB-IDLMA provided higher performance for Vo./Dr. and similar performance for the other instrument pairs.
These results show that the proposed EB-IDLMA can achieve the best performance for all instrument pairs without the adequate choice of $\nu$, which is required in $t$-IDLMA.
Furthermore, the best $\nu$ value of $t$-IDLMA varied with songs, as shown in Table~\ref{tab:ratio_nu_tIDLMA}.
Since EB-IDLMA can output different $\nu_{ijn}$ for each song, the proposed direction would also be promising for handling this variation, which we leave as our future work.

\vspace{-2mm}
\mysection{CONCLUSION}
\label{sec:conclusion}
We proposed EB-IDLMA by extending IDLMA to encompass the reliability of the source power spectrograms estimated by DNNs.
The key idea of EB-IDLMA is that DNNs estimate not the scale parameters but the hyperparameters of the prior distribution of the scale parameters.
These hyperparameters determine the reliability of the estimated source power spectrograms, and in the demixing matrix estimation, they were updated by DNNs trained on the basis of the empirical Bayes method.
Experimental evaluations show the effectiveness of EB-IDLMA and the importance of considering the reliability of the source power spectrograms estimated by DNNs.

\vspace{-0.5mm}
\appendix
\def\thesection{\arabic{mysection}}
\setcounter{section}{5}
\mysubsection{Proof of Proposition~1}
\label{sec:appendix/proof1}
When $|\tilde{s}_{ijn}|^2\ll \delta$ and $\hat{r}_{ijn}^2\ll \delta$, the cost function $\mathcal{L}_{\mathrm{EB}}^{(\DNN_{n})}$ approximately reduces to 
\begin{equation}
    \mathcal{L}_{\mathrm{EB}}^{(\DNN_{n})} \simeq \sum_{i,j}
    \left[
        \log\delta+
        {\left(1+\cfrac{\hat{\nu}_{ijn}}{2}\right)}\log{\left(1+\cfrac{2}{\hat{\nu}_{ijn}}\right)}
    \right].
    \label{eq:L_approx}
\end{equation}
The derivative of the right-hand side of \eqref{eq:L_approx} w.r.t. $\hat{\nu}_{ij}$ is described as
\begin{align}
    \frac{\partial \mathcal{L}_{\mathrm{EB}}^{(\DNN_{n})}}{\partial \hat{\nu}_{ij}}
    =& \frac{1}{2}\log\left(1+\frac{2}{\hat{\nu}_{ijn}}\right)-\frac{1}{\hat{\nu}_{ijn}}.
    \label{eq:experimental-evaluation/dof-estimation}
\end{align}
To examine the sign of this derivative, we consider the function $f(\eta)=\ln(1+2\eta)/2$ of $\eta>-1/2$.
Owing to its concavity, the tangent line at $\eta=0$ is greater than $f(\eta)$ at any $\eta>0$: $\log(1+2\eta)/2 < \eta$.
By replacing $\eta$ with $1/\nu_{ijn}$, we can confirm that the right-hand side of \eqref{eq:experimental-evaluation/dof-estimation} is below zero for any $\hat{\nu}_{ijn}$.

\bibliographystyle{IEEEtran}
\bibliography{refs_v10}

\end{document}